\documentclass[12pt,a4paper]{article}
\usepackage{amsmath}
\usepackage{graphicx}
\pdfoutput=1
\usepackage{times}
\begin{document}
\begin{titlepage}
\title{Comment on the ``extended eikonal'' unitarization}
\author{ S.M. Troshin\\[1ex]
\small  \it Institute for High Energy Physics,\\
\small  \it Protvino, Moscow Region, 142281, Russia}
\normalsize
\date{}
\maketitle

\begin{abstract}
This is a comment on the recent paper by O.V.~Selyugin, J.-R.~Cudell,
and E.~Predazzi "Analytic properties of different
unitarization schemes" arXiv: 0712.0621v2, [hep-ph].
\end{abstract}
\end{titlepage}
\setcounter{page}{2}

Unitarity or conservation of probability, which can be   written
in terms of the scattering matrix as
\begin{equation}\label{ss}
SS^+=1,
\end{equation} implies an
existence at high energies of the two scattering modes - shadowing
and antishadowing. Antishadowing  in the context of the rising
total cross-sections and transition beyond the black disk limit
was discussed in the framework of the rational unitarization
scheme in \cite{phl} (Fig. 1).
\begin{figure}[h]
\resizebox{16cm}{!}{\includegraphics*{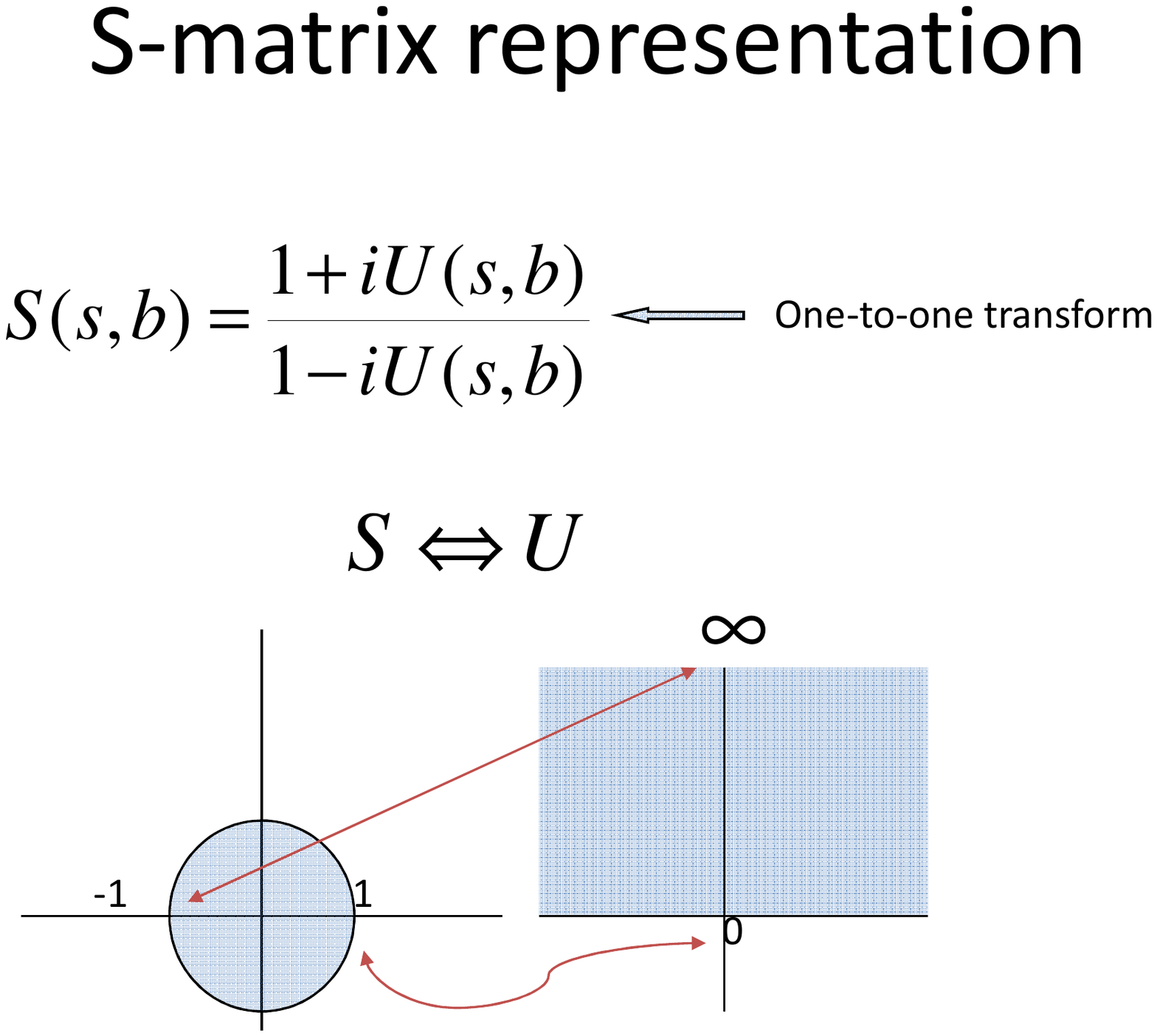}}
\caption{Rational unitarization scheme (color online)}
\end{figure}
In the recent paper \cite{reflect} we considered saturation of the
unitarity limit in head-on (and small impact parameter) hadron
collisions, i.e. when $S(s,b)|_{b=0}\to -1$ at $s\to\infty$. It
was shown that the approach to the full absorption in head-on
collisions, in another words, the limit $S(s,b)|_{b=0}\to 0$ at
$s\to\infty$ does not follow from unitarity, it is merely  a
result of the assumed saturation of the black disk limit. This
limit is a direct consequence of the exponential unitarization
with an extra assumption on the pure imaginary nature of the phase
shift (Fig. 2).
\begin{figure}[h]
\begin{center}
\resizebox{16cm}{!}{\includegraphics*{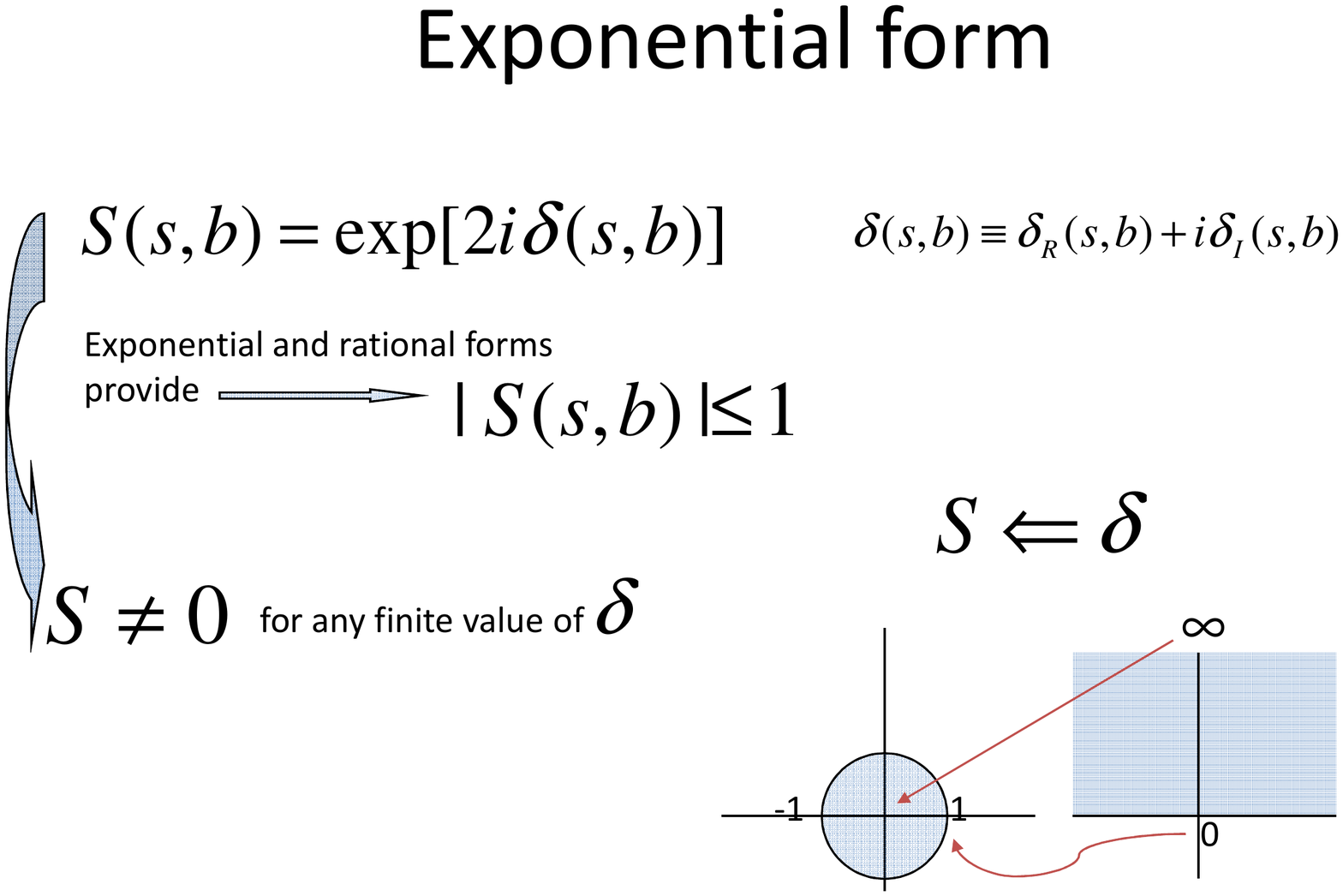}}
\end{center}
\caption{Exponential unitarization scheme (color online)}
\end{figure}
Natural interpretation  of antishadowing as the
the reflective scattering was proposed. This interpretation of the
unitarity saturation has grounds in the optical concepts in high energy
hadron scattering.
A new attempt to derive antishadowing scattering mode was made in
\cite{sel}. The authors claim that ``extended eikonal''
unitarization scheme can also reproduce antishadowing.
\begin{figure}[h]
\begin{center}
\resizebox{16cm}{!}{\includegraphics*{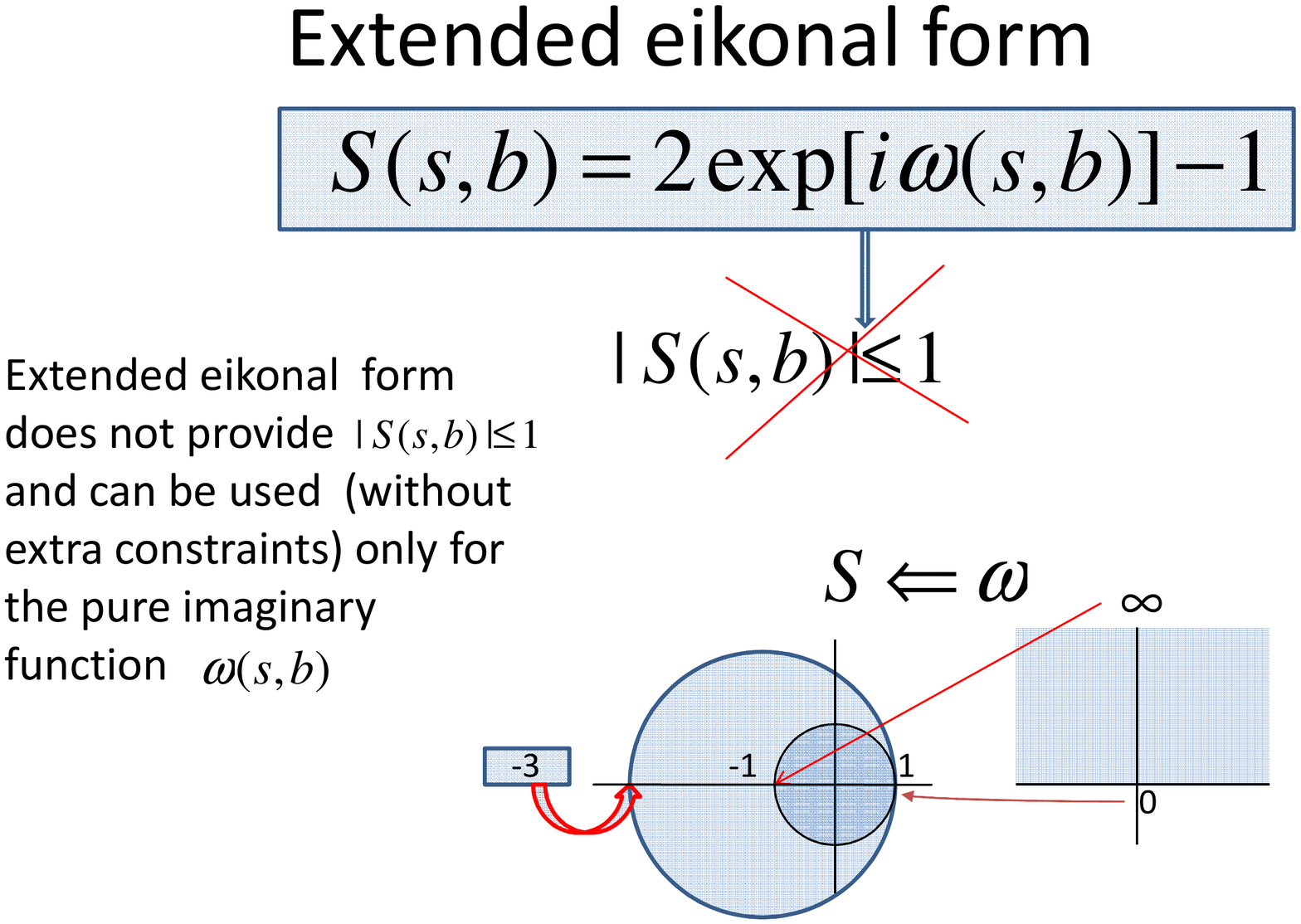}}
\end{center}
\caption{Extended eikonal unitarization scheme (color online)}
\end{figure}
However, as it is evident from Fig. 3 the extended eikonal itself
cannot be considered as a unitarization scheme.
The authors of \cite{sel} in the new version of their paper consider
possible unitarity violation when a Born amplitude has a real part
and claim that such violation is small at high energies. This argument
does not work at large values of impact parameters where the eikonal function
is vanishingly small.
\section*{Acknowledgement}
I am grateful to Laszlo Jenkovszky and Oleg Selyugin for the
interesting and extensive correspondence.

\end{document}